\documentclass[doublecol]{epl2} 
\usepackage{graphicx,rotating,subfigure,amsmath,amsfonts,amssymb,delarray,epstopdf}

\newcommand{\e}{\text{e}}
\newcommand{\im}{\text{i}}
\def\l{\left}
\def\r{\right}
\def\12{\frac{1}{2}}

\title{NMR Response in quasi one-dimensional Spin-$\frac{1}{2}$ Antiferromagnets}
\author{J. Sirker\inst{1} \and N.Laflorencie\inst{2,3,4}}
\shortauthor{J. Sirker \and N.Laflorencie} % \etal}

\institute{                    
  \inst{1} Max-Planck-Institut f\"ur Festk\"orperforschung -
              Heisenbergstrasse 1, D-70569 Stuttgart, Germany\\
  \inst{2} Laboratoire de Physique des Solides, Universit\'e Paris-Sud - 
  UMR-8502 CNRS, 91405 Orsay, France\\
  \inst{3} Universit\'e de Toulouse; UPS; Laboratoire de
Physique Th\'eorique (IRSAMC) - F-31062 Toulouse, France\\
\inst{4} CNRS; LPT (IRSAMC) - F-31062 Toulouse, France}

\abstract{Non-magnetic impurities break a quantum spin chain into finite
  segments and induce Friedel-like oscillations in the local susceptibility
  near the edges.  The signature of these oscillations has been observed in
  Knight shift experiments on the high-temperature superconductor
  YBa$_2$Cu$_3$O$_{6.5}$ and on the spin-chain compound Sr$_2$CuO$_3$. Here we
  analytically calculate NMR spectra, compare with the available experimental
  data for Sr$_2$CuO$_3$, and show that the interchain coupling is responsible
  for the complicated and so far unexplained lineshape. Our results are based
  on a parameter-free formula for the local susceptibility of a finite spin
  chain obtained by bosonization which is checked by comparing with quantum
  Monte Carlo and density-matrix renormalization group calculations.}

\begin{document}
\bibliographystyle{eplbib}
\maketitle

\section{Introduction}
An important tool to study the {\it local} spin dynamics in strongly
correlated electron systems is nuclear magnetic resonance (NMR).  NMR
experiments have been instrumental in investigating spin fluctuations and
impurity effects in high-temperature superconductors
\cite{TakigawaReyesAlloulMendels}, as well as in confirming the triplet nature
of superconductivity in Sr$_2$RuO$_4$ \cite{IshidaMukuda}. Quite recently, NMR
was also used to study the CuO chains in YBa$_2$Cu$_3$O$_{6.5}$ (YBCO)
\cite{YamaniStatt}.  The NMR study showed that the chain ends induce
Friedel-like oscillations which manifest themselves also in the CuO$_2$
planes. Similar oscillations have also been observed earlier in the
prototypical quasi one-dimensional $S=1/2$ spin chain compound Sr$_2$CuO$_3$
(SCO) \cite{TakigawaMotoyama,BoucherTakigawa}.  Theoretically, a large
alternating component of the local susceptibility near the end of a
semi-infinite Heisenberg chain has been predicted \cite{EggertAffleck95}.
Other studies (for a recent review see ref.~\cite{AlloulBobroff}) have
addressed local spin correlations near a chain end by numerical means in a
variety of one-dimensional models ranging from the frustrated and dimerized
spin-$1/2$ chain to spin ladders and the spin-$1$ Heisenberg
chain \cite{OtherStudies}.
% ~\cite{MartinsLaukamp,LaukampMartins, Tedoldi,AletSoerensen,DasMahajan}.

An interesting open question concerning the physics of the spin-$1/2$
Heisenberg chain is whether its transport properties are ballistic or
diffusive. The theoretical results are contradictory
\cite{Zotos,KluemperScheeren,Heidrich-Meisner,SirkerDiff,Giamarchi,RoschAndrei}
but seem to point to ballistic transport perhaps related to the integrability
of the model by Bethe ansatz.  NMR and muon spin relaxation experiments on SCO
\cite{Thurber,muSR}, on the other hand, have found diffusive behavior. In
order to analyze these experiments on a quantitative level in the future, it
is first of all important to know in how far a spin-only model for SCO is
valid. In particular, it has been claimed in ref.~\cite{BoucherTakigawa} that
already the NMR spectra - where only static correlations are tested - can only
be understood if a coupling to the lattice is taken into account.

In the first part of this letter we will calculate NMR spectra for a
Heisenberg chain with a Poisson distribution of non-magnetic impurities.This
is known to be the relevant model for SCO, with chain breaks caused by the
presence of excess oxygen \cite{MotoyamaEisaki,SirkerLaflorencie}. We will
start with the ideal chain but will then show that the interchain couplings
are essential to fully explain the experimental data for SCO
\cite{TakigawaMotoyama,BoucherTakigawa}.  Our main findings are that the
theoretically calculated spectra for a spin-only model of weakly coupled
Heisenberg chains are in perfect agreement with experiment whereas the
phenomenological model proposed in \cite{BoucherTakigawa}, involving some
coupling to lattice degrees of freedom, cannot - if taken seriously - explain
the data.
%We will show that the experimental data for SCO
%\cite{TakigawaMotoyama,BoucherTakigawa} can be fully explained if the
%interchain coupling is taken into account and that the model of mobile defects
%proposed in \cite{BoucherTakigawa} is incorrect. 
The calculated NMR spectra are based on a {\it parameter-free} formula for the
local susceptibility of {\it finite} $S=1/2$ $XXZ$ chains at finite
temperatures
% including the logarithmic corrections at the isotropic point 
obtained by a bosonization approach as explained in the second part of this
letter. The analytical results allow for a full impurity averaging which would
be impossible to achieve at low temperatures by numerical calculations.
\begin{figure}[ht!]
\includegraphics*[width=\columnwidth]{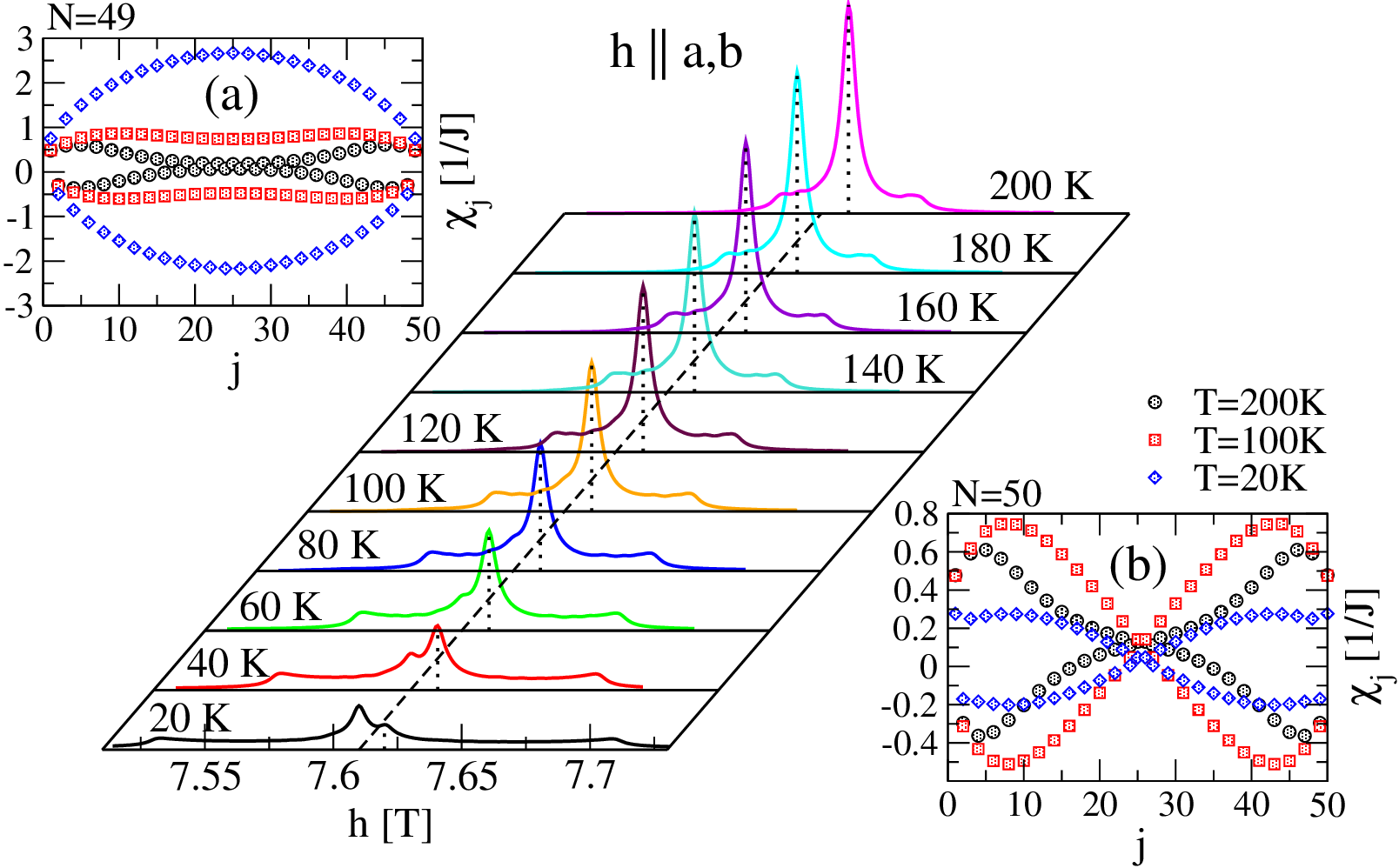}
\caption{(Color online) Theoretical NMR spectra (\ref{Ks2}) for a Poisson
  distribution with $p=0.3\%$ non-magnetic impurities, parameters $J,\, A^0,\,
  A^{\pm 1}$ as appropriate for SCO (see text), $\Gamma=4\cdot 10^{-4}$, and
  $h_{\rm res}^0 = 7.61$ T (dashed line). The dotted lines mark the maxima
  corresponding to the bulk susceptibility.  Insets (a) and (b) show local
  susceptibilities %for odd and even lengths
  at various temperatures indicated on the plot.}
\label{fig1}
\end{figure}
\section{NMR spectra}
The Hamiltonian of the spin-$1/2$ $XXZ$ model with $N$ sites and open boundary
conditions (OBCs) is given by
\begin{equation}
\label{eq1}
H = J\sum_{j=1}^{N-1} \l[ S^x_j S^x_{j+1} + S^y_j S^y_{j+1} + \Delta S^z_j
S^z_{j+1} \r] - h\sum_{j=1}^N S^z_j \; .
\end{equation}
Here $J$ is the exchange constant, $\Delta\in [0,1]$ an exchange
anisotropy, and $h$ the applied magnetic field. Due to the OBCs,
translational invariance is broken leading to a position dependent local 
susceptibility
 \begin{equation}
\label{eq2} 
\chi^{(N)}_j=\frac{\partial}{\partial h} \langle S^z_j\rangle_{h=0} = \frac{1}{T} \langle
S^z_j S^z_{\rm tot}\rangle_{h=0}
\end{equation}
where $T$ is the temperature and $S^z_{\rm tot} = \sum_j S^z_j$. The hyperfine
interaction couples nuclear and electron spins. For a chain segment of length
$N$ this leads to the {\it Knight shift} of the nuclear resonance frequency
$K^{(N)}_j=(\gamma_e/\gamma_n)\sum_{j'}A^{j-j'}\chi^{(N)}_{j'}$, where
$\gamma_e$ ($\gamma_n$) is the electron (nuclear) gyromagnetic ratio,
respectively. The hyperfine interaction is short ranged so that usually only
$A^0$ and $A^{\pm 1}$ matter. The NMR spectrum is proportional to the
distribution of Knight shifts. Let us assume in the following a Poisson
distribution of non-magnetic impurities with concentration $p$ and a
Lorentzian lineshape with width $\Gamma$ for each Knight shift. The normalized
probability distribution is then given by
\begin{equation}
\label{Ks2}
 P(K) = \frac{\Gamma}{\pi}\!\!\sum_{N=1}^\infty
\frac{p(1-p)^{N-1}}{N}\!\sum_{j=1}^N\frac{1}{(K-K^{(N)}_j)^2+\Gamma^2} \; .
\end{equation}
As we will show in the second part of this letter, bosonization allows us to
derive a {\it parameter-free} result for $\chi_j^{(N)}$ in the limit $T/J\ll
1$ and $N\gg 1$. Because the deviations for very small chain lengths are not
important for the NMR spectra as long as the probability of having such tiny
segments is low, the only parameters entering in (\ref{Ks2}) are the
material-dependent constants $J$, $A^0$ and $A^{\pm 1}$.

NMR measurements have been performed on the Heisenberg ($\Delta=1$) chain
compound SCO \cite{TakigawaMotoyama,BoucherTakigawa}. Chain breaks in this
system are believed to be caused by randomly distributed excess oxygen leading
to the formation of Zhang-Rice singlets
\cite{MotoyamaEisaki,SirkerLaflorencie}.  From measurements of the total
susceptibility it follows that $J\sim 2200$ K \cite{MotoyamaEisaki}. By a
comparison with YBa$_2$Cu$_3$O$_{6+\delta}$ \cite{MonienPines} and theory the
hyperfine coupling constants $A_c^0/(2\hbar\gamma_n) \approx -13$ T,
$A_{ab}^0/(2\hbar\gamma_n) \approx 2$ T, and $A^1/(2\hbar\gamma_n) \approx 4$
T are obtained. Here the index denotes the magnetic field direction. We
calculate the spectra as a function of $h=(1+K)h_{\rm res}^0$ where $h_{\rm
  res}^0=\nu/\gamma_n$ is the resonance field for an isolated $^{63}$Cu atom.
In experiment $\nu=86$ MHz \cite{TakigawaMotoyama} and $\gamma_n\approx 11.3$
MHz/T \cite{AbragamBleaney} leading to $h_{\rm res}^0\approx 7.61$ T.
Exemplarily, we show the evolution of the lineshape for an ideal chain with
impurity concentration $p=0.3\%$ in fig.~\ref{fig1}.
% In fig.~\ref{fig1}, the spectra for $h\parallel a,b$ and an impurity
% concentration $p=0.3\%$ are shown.  They are dominated
At high temperatures a central peak dominates whose position corresponds to
the bulk susceptibility value. In addition, broad edges are visible whose
separation increases $\sim h^0_{\rm res}\sqrt{v/T}\ln^{1/4}(v/T)$ with
decreasing temperature with $v$ being the spin velocity. These edges are
caused by the extrema in the local susceptibilities of chain segments with
lengths $N\gg \pi v/T$ (see Figs.~\ref{fig1} (a) and (b), respectively).
Furthermore, we observe a gradual transfer of weight from a peak at high
temperatures to a peak corresponding to zero Knight shift at low temperatures
stemming from the increasing number of even chain segments with $N\ll \pi v/T$
which become frozen into their singlet ground state (see fig.~\ref{fig1}(b)).
The odd chain segments with $N\ll \pi v/T$, on the other hand, will yield large
Knight shifts (see fig.~\ref{fig1}(a)). This leads to a background which grows
in intensity and expands with decreasing temperature. The temperature $T^*$
where the peak corresponding to zero Knight and the peak corresponding to the
bulk susceptibility value have equal height is therefore directly related to
the impurity concentration. For $p\ll 1$ we analytically find
\begin{equation}
  \label{crit}
  p\approx 1-3^{-2T^*/(J\pi^2)} 
  \end{equation} 
  which is a simple criterion to determine the impurity concentration from the
  NMR spectra alone.

\begin{figure}[t!]
\includegraphics*[width=\columnwidth]{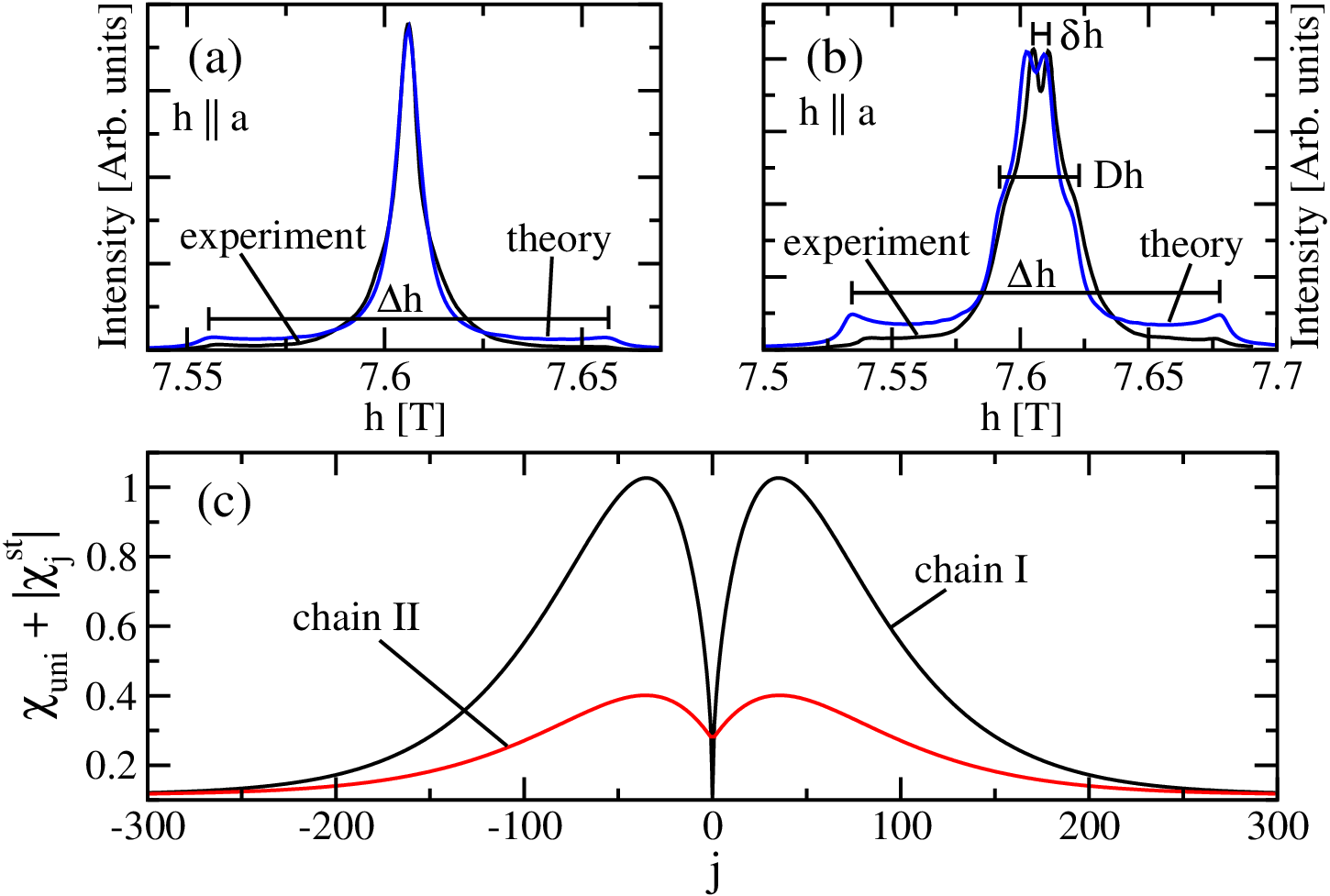}
\caption{(Color online) (a) Comparison between experimental data taken from
  \cite{TakigawaMotoyama} and theory with material-dependent parameters as
  given in the text and $p=5\times 10^{-4}$, $\Gamma=4\times 10^{-4}$ at $T=60$
  K. The peak height has been adjusted to agree with experiment and $h^0_{\rm
    res}=7.5955$. (b) Same as (a) at $T=30$ K. (c) Calculated oscillations $\chi_j$
  at $T=30$ K in an infinite chain with an impurity at $x=0$ (chain I) and
  reflections in the infinite chain II without impurities due to an interchain
  coupling with strength $J_\perp =5$ K.}
\label{fig2}
\end{figure}

  The observation of a central peak at high temperatures and broad edges
  with separation $\Delta h\sim h^0_{\rm res}\sqrt{v/T}$ is in agreement with
  experimental observations \cite{TakigawaMotoyama,BoucherTakigawa} as shown
  in fig.~\ref{fig2}. It has indeed already been pointed out in
  \cite{TakigawaMotoyama} - based on the theoretical results for the
  semi-infinite chain by Eggert and Affleck \cite{EggertAffleck95} - that the
  edges are a consequence of the maxima in the local susceptibility.
However, at temperatures $T\lesssim 30$ K additional structures are visible in
the experimental spectra.  The peak develops shoulders whose separation is
denoted by $Dh$ in fig.~\ref{fig2}(b) following the notation introduced in
\cite{TakigawaMotoyama}. Furthermore, a splitting of the peak, $\delta h$,
very different from the weight transfer with temperature shown in
fig.~\ref{fig1}, is observed.  In ref.~\cite{BoucherTakigawa} it has been
tried to explain these features by a phenomenological model of mobile bond
defects. In particular, the feature $\delta h$ was ascribed to a periodic
arrangement of bond defects leading to chain segments of odd length $N$ only.
However, analytically we find that the splitting of the central peak would
then grow like $\sim h^0_{\rm res} J\ln^{1/4}(N)/(TN^{1/2})$. Due to the
prefactor $\sim J$ this predicts a very rapid increase of the splitting not
observed in experiment
%making this model implausible. 
showing that this model is incorrect. Instead, as we will show below, the
additional features $Dh$ and $\delta h$ are a consequence of the interchain
coupling. The interchain coupling along one of the crystal axes perpendicular
to the chains is of order $J_\perp\sim 5$ K while it is three orders of
magnitude smaller along the other direction \cite{MotoyamaEisaki,Rosner}. At
$T\sim 30$ K we might therefore already expect significant effects of the
interchain couplings which, however, can be included perturbatively. Doing so
we find that the susceptibility oscillations near an impurity residing in one
chain (zeroth order) lead to substantial reflections in the neighboring chain
(first order) for parameters appropriate for SCO as shown in
fig.~\ref{fig2}(c). The ratio of the maximum in chain I to the maximum in
chain II is $\sim \Delta h/(Dh)$, i.e., the shoulders of the peak, $Dh$, are
caused by the maxima of the reflected oscillations. We find $Dh\sim h^0_{\rm
  res} J_\perp\sqrt{v} T^{-3/2}$ but logarithmic corrections can disguise this
scaling as will become clear later on. The splitting of the central peak
$\delta h$, see fig.~\ref{fig2}(b), has a more complicated origin. First,
there are also a reflections in next-nearest neighboring chains (second
order). The maxima would then yield a splitting $\delta h\sim h^0_{\rm res}
J_\perp^2\sqrt{v} T^{-5/2}$ (again ignoring logarithmic corrections). However,
for impurity concentrations $p\sim 5\times 10^{-4}$ relevant for the
experiments there is another effect which actually dominates: Including the
first and second order reflections from neighboring chains, the chain of
average length $\bar{N}\approx 1/p$ will not have any sites left which show
bulk behavior.  This means that at temperatures $T\lesssim 30$ K the
probability for having Knight shifts corresponding to values close to the bulk
susceptibility starts to decrease dramatically thus leading to a drop in
intensity in $P(K)$. Therefore $\delta h$ denotes not a splitting of the
central peak but rather a dip of intensity at the bulk susceptibility value.
The oscillations, which now basically spread over the entire crystal, might
get further stabilized by anisotropic exchange terms. This might also explain
the small differences in the lineshape near the peak for $h\parallel a$ and
$h\parallel c$ \cite{TakigawaMotoyama,BoucherTakigawa}. A detailed analysis of
these anisotropy effects is beyond the scope of this letter.

We also want to stress that having additional structures in the spectra due to
interchain couplings is very different from a scenario where instead such
structures occur due to a direct hyperfine interaction or a dipolar coupling
of the nuclear spins with electron spins in adjacent chains. The latter case
would lead to $Dh/\Delta h =\mbox{const.}$ ($\delta h/\Delta h =
\mbox{const.}$) as a function of temperature if the shoulders (splitting) are
caused by this mechanism, respectively. Only three (two) different
temperatures are presented in \cite{BoucherTakigawa} where the shoulders
(splitting) are visible, respectively, making a detailed analysis
impossible. However, we find that the data are consistent with $Dh/\Delta
h\sim 1/T$ as expected if interchain coupling dominates but certainly not
consistent with either $Dh/\Delta h =\mbox{const.}$ or $\delta h/\Delta h
=\mbox{const.}$.

To calculate the NMR lineshapes for SCO at low temperatures, we have
to deal with a two-dimensional array of weakly coupled chains. In
general, this is a very complicated task requiring a two-dimensional
impurity averaging.  However, for small impurity concentrations
significant simplifications are possible. At a given temperature, the
oscillations extend over a characteristic length $\xi$.  If
$\bar{N}=1/p\gg \xi$ then the probability of having two impurities in
neighboring chains so close to each other that the zeroth order
oscillations in the chain and the reflections from neighboring chains
overlap is small.  We therefore assume that reflections in a chain of
length $N$ only occur in regions where the chain shows bulk
behavior. In a chain of length $N$, $2pN$ reflections from the nearest
neighbor chains and $2pN$ reflections from nearest-neighbor chains
will occur on average. If a chain segment is long enough, we consider
the zeroth, first, and second order oscillations as independent
entities. If the segment is too short, we reduce the extend of the
first and second order oscillations mimicking the overlap.  For the
chain segments we then do the full impurity averaging (\ref{Ks2}). The
theoretically calculated lineshapes we obtain this way are in
excellent agreement with experiment as shown in fig.~\ref{fig2}. For
both temperatures we use the peak to adjust the intensity. For $T=60$
K we also used $p$ and $h^0_{\rm res}$ as parameters and find
$p=5\times10^{-4}$ and $h^0_{\rm res}=7.5955$ T.  In fig.~\ref{fig2}
(b) we use the same values. An even better agreement would be obtained
here if we choose $h^0_{\rm res}=7.598$ T which is equivalent to a
deviation of $0.03\%$ from our prediction for the evolution of the
bulk susceptibility value.  Furthermore, the intensity of the
edges $\Delta h$ is overestimated.  This is most likely a
consequence of our assumption that the zeroth order oscillations do
not overlap with the reflections. Configurations where such an overlap
occur would wash out the maxima of the zeroth order oscillations. In
addition, this might also point to some deviations from a Poisson
distribution with short chains occurring less frequently than expected.

\section{Local susceptibility}
We now explain how the parameter-free results for $\chi_j^{(N)}$ have been
obtained. In the low-energy limit, the spin operators can be expressed in
terms of a boson $\Phi$ as
\begin{equation}
\label{eq3}
S^z_j \approx \sqrt{\frac{K_L}{2\pi}}\partial_x\Phi + c (-1)^j \cos\sqrt{2\pi
    K_L}\Phi \; .
\end{equation}
Here $K_L$ is the Luttinger parameter and $c$ the amplitude of the alternating
part. The integrability of model (\ref{eq1}) by Bethe ansatz allows it to
determine $K_L$ and $c$ exactly for all $\Delta$. Ignoring bulk and boundary
irrelevant operators, the Hamiltonian (\ref{eq1}) is equivalent to a free
boson model
\begin{eqnarray}
\label{eq4}
H \!&=& \!\frac{v}{2} \int_0^{L+a} \!\!\!\!\!\!\!\! dx \l[\Pi^2+(\partial_x\Phi)^2\r] -
h\sqrt{\frac{K_L}{2\pi}}\!\int_0^{L+a} \!\!\!\!\!\!\!\! dx \,\partial_x\Phi %% \nonumber \\
\end{eqnarray}
where $v$ is the spin velocity, $L=Na$, and $a$ the lattice constant.  The
bosonic fields obey the standard commutation rule
$[\Phi(x),\Pi(x')]=\im\delta(x-x')$ with $\Pi=v^{-1}\partial_t\Phi$. To
calculate the local susceptibility (\ref{eq2}) we use a mode expansion
\begin{eqnarray}
\label{ModeExp}
&& \Phi(x=ja,t) = \sqrt{\frac{\pi}{8K_L}} +\sqrt{\frac{2\pi}{K_L}} S^z_{\rm tot} \frac{j}{N+1} \\
&+& \sum_{n=1}^\infty
\frac{\sin\l(\pi n j/(N+1)\r)}{\sqrt{\pi n}}\l(\e^{-i\frac{\pi n vt}{L+a}}b_n +
\e^{i\frac{\pi n vt}{L+a}}b_n^\dagger\r) \nonumber
\end{eqnarray}
which incorporates the OBCs. Here $b_n$ is a bosonic annihilation operator.
Eq.~(\ref{ModeExp}) is a discrete version of the mode expansions used in
\cite{EggertAffleck92,SirkerLaflorencie} with $x=ja$ becoming a continuous
coordinate for $a\to 0$, $N\to\infty$ with $L=Na$ fixed. Using this mode
expansion, the local observables respect the discrete lattice symmetry $j\to
N+1-j$ corresponding to a reflection at the central bond (site) for $N$ even
(odd), respectively. The sites $0$ and $N+1$ are added to model (\ref{eq1})
and we demand that the spin density vanishes at these sites. Therefore the
upper boundary for the integrals in (\ref{eq4}) is $L+a$. The zero mode part
(first line of eq.~(\ref{ModeExp})) fulfills $\sum_j S^z_j \approx
\sqrt{\frac{K_L}{2\pi}}\int_0^{L+a} \partial_x\Phi \equiv S^z_{\rm tot}$ and the
oscillator part (second line of eq.~(\ref{ModeExp})) vanishes for $j=0,\, N+1$
as required.

Using (\ref{eq3}) in the formula for the local susceptibility (\ref{eq2}), we
find $\chi_j =\chi_j^{\rm uni} + (-1)^j\chi_j^{\rm st}$. The uniform part is
independent of position, $\chi_j^{\rm uni}\equiv \chi^{\rm uni} = \langle
(S^z_{\rm tot})^2\rangle/(TN)$ and we can therefore directly use the
parameter-free result derived in \cite{SirkerLaflorencie}.
\begin{figure}
\includegraphics*[width=\columnwidth]{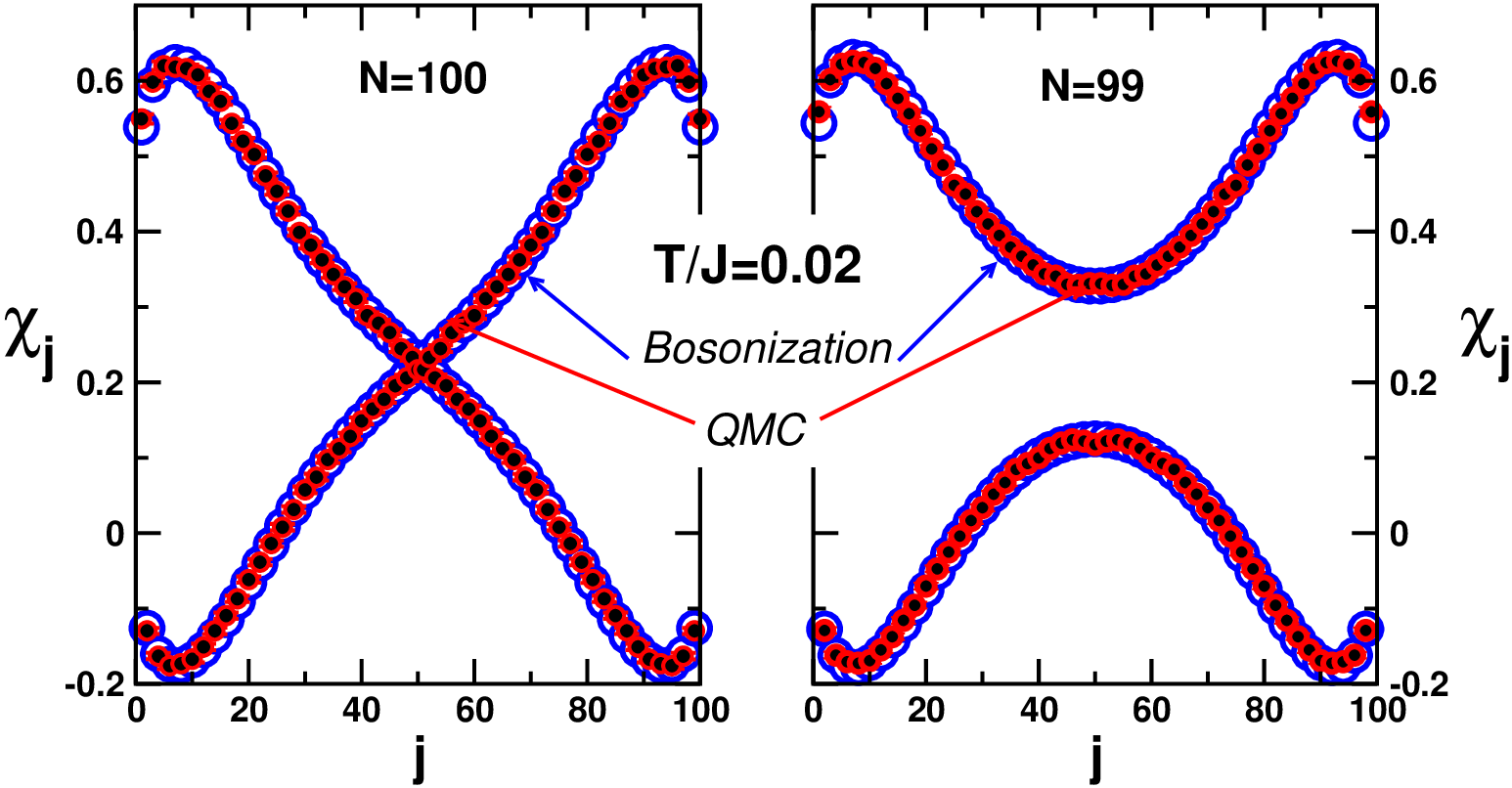}
\caption{(Color online) $\chi_j$ for $\Delta=0.3$ and $T/J=0.02$. Comparison between QMC and field theory for $N=100$ and $N=99$.}
\label{fig3}
\end{figure}
For the staggered part, on the other hand, we find $ \chi_j^{\rm st} =
\frac{c}{T}\langle\cos\sqrt{2\pi K_L}\Phi\rangle_{\rm osc}\langle\cos\sqrt{2\pi
  K_L}\Phi S^z_{\rm tot}\rangle_{\rm zm} $ where we have split the correlation
function into an oscillator and a zero mode part according to (\ref{ModeExp}).
Using the cumulant theorem for bosonic modes $\langle\exp(\pm \im\sqrt{2\pi
  K_L}\Phi)\rangle_{\rm osc} = \exp(-\pi K_L\langle\phi\phi\rangle_{\rm osc})$ we
obtain, following \cite{MattsonHortonEggert,SirkerLaflorencie},
\begin{equation}
\label{StaggPart2}
\langle\cos\sqrt{2\pi K_L}\Phi\rangle_{\rm osc}
=\l(\frac{\pi}{N+1}\r)^{K_L/2}\!\!\!\!\!\!\frac{\eta^{3K_L/2}\l(\e^{-\frac{\pi
    v}{TL}}\r)}{\theta_1^{K_L/2}\l(\frac{\pi j}{N+1},\e^{-\frac{\pi
    v}{2TL}}\r)} \, .
\end{equation}
Here $\eta(x)$ is the Dedekind eta-function and $\theta_1(u,q)$ the elliptic
theta-function of the first kind. % Furthermore,
For the zero mode part we find
\begin{eqnarray}
\label{StaggPart3}
&& \langle\cos\sqrt{2\pi K_L}\Phi S^z_{\rm tot}\rangle_{\rm zm} \\
&=& -\frac{\sum_{m} m\sin[2\pi m j/(N+1)]\e^{-\pi vm^2/(K_LLT)}}{\sum_{m}\e^{-\pi
    vm^2/(K_LLT)}} \nonumber
\end{eqnarray}
with $m$ running over all integers (half-integers) for $N$ even (odd),
respectively. In the thermodynamic limit, $N\to\infty$, we can simplify our
result and obtain
\begin{equation}
\label{StaggPart4}
\chi_j^{\rm st} = \frac{cK_L}{v}\frac{x}{\l[\frac{v}{\pi T}\sinh\l(\frac{2\pi T x}{v}\r)\r]^{K/2}}
\end{equation}
with $x=ja$. This agrees for the isotropic Heisenberg case, $K_L=1$, with the
result in \cite{EggertAffleck95}.  The amplitude $c$ % in eq.~(\ref{eq3})
can be determined with the help of the Bethe ansatz along the lines of
ref.~\cite{LukyanovTerras}.  This leads to $c=\sqrt{A_z/2}$ with $A_z$ as
given in eq.~(4.3) of \cite{LukyanovTerras}.  Our result for the staggered
part of the local susceptibility is therefore {\it parameter free}.  This
means that we can directly compare our analytical result for $\chi_j$ with
quantum Monte Carlo (QMC) data.  For an anisotropy $\Delta=0.3$, shown in
fig.~\ref{fig3}, the agreement is excellent.
\begin{figure}
\includegraphics*[width=\columnwidth]{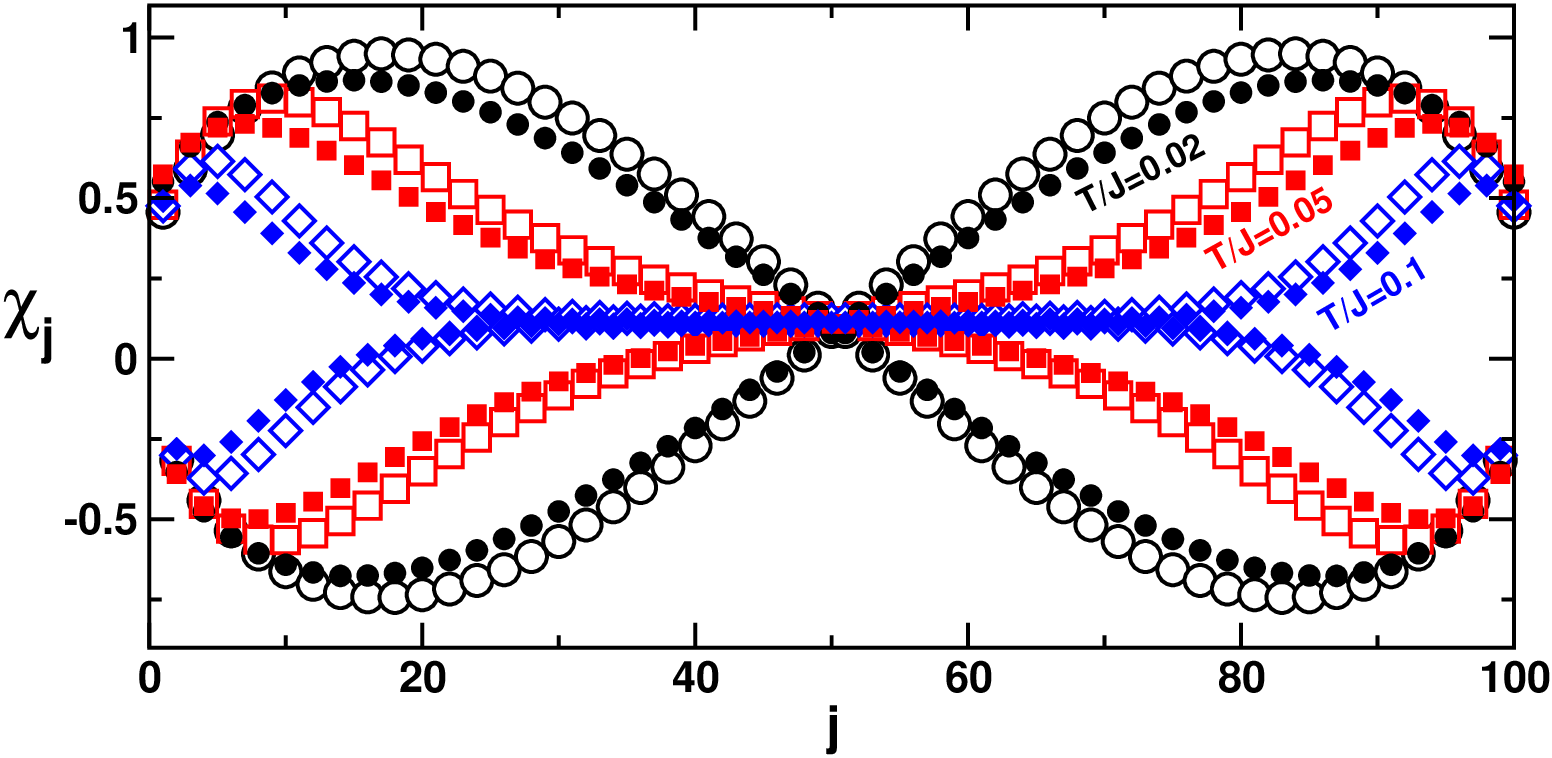}
\caption{(Color online) $\chi_j$ for $\Delta=1$ and $N=100$. Comparison
  between QMC (closed symbols) and field theory (open symbols) for
  $T/J=0.02,~0.05,~0.1$, as indicated on the plot.}
\label{fig4}
\end{figure}

Next, we come to the experimentally most relevant isotropic case, $\Delta=1$
($K_L=1$). Umklapp scattering is then marginally irrelevant and the scaling
dimensions of correlation functions have to be replaced by renormalization
group improved versions. For $\chi_j$ the calculations are rather similar to
those in ref.~\cite{aff_corr} for the longitudinal spin-spin correlation
function. We find that we have to replace $1/K_L\to 1-g$ in (\ref{StaggPart3})
whereas $K_L\equiv 1$ in the oscillator part, eq.~(\ref{StaggPart2}). The
renormalization of $K_L$ for this part is incorporated into an effective
amplitude $c\to (2\pi^3 g)^{-1/4}$. % in eq.~(\ref{StaggPart}).
The running coupling constant $g$ depends, in general, on the three length
scales $x$, $L$, and $v/T$. At low enough energies the smallest scale will
always dominate and $g$ is given by the solution of $1/g +\ln(g)/2 =\ln\l\{C_0
\min[x,L-x,v/T]\r\}$ where $C_0$ is a constant. In fig.~\ref{fig4} a
comparison between this analytic result and QMC data is shown with $C_0=6$. We
note that fitting the constant $C_0$ improves the results near the boundaries.
For low temperatures and $x,L-x\gg 1$, however, the value of $C_0$ becomes
irrelevant and our result for $\chi_j$ therefore again parameter-free.  The
agreement with QMC is not as good as for the anisotropic case. This is a
consequence of the fact that $g$ has been derived in the limit $x,L,v/T\gg1$.
However, the deviations are only of the order of a few percent and have very
little effect on the NMR spectra presented in the first part of this
letter.

\begin{figure}
\includegraphics*[width=\columnwidth]{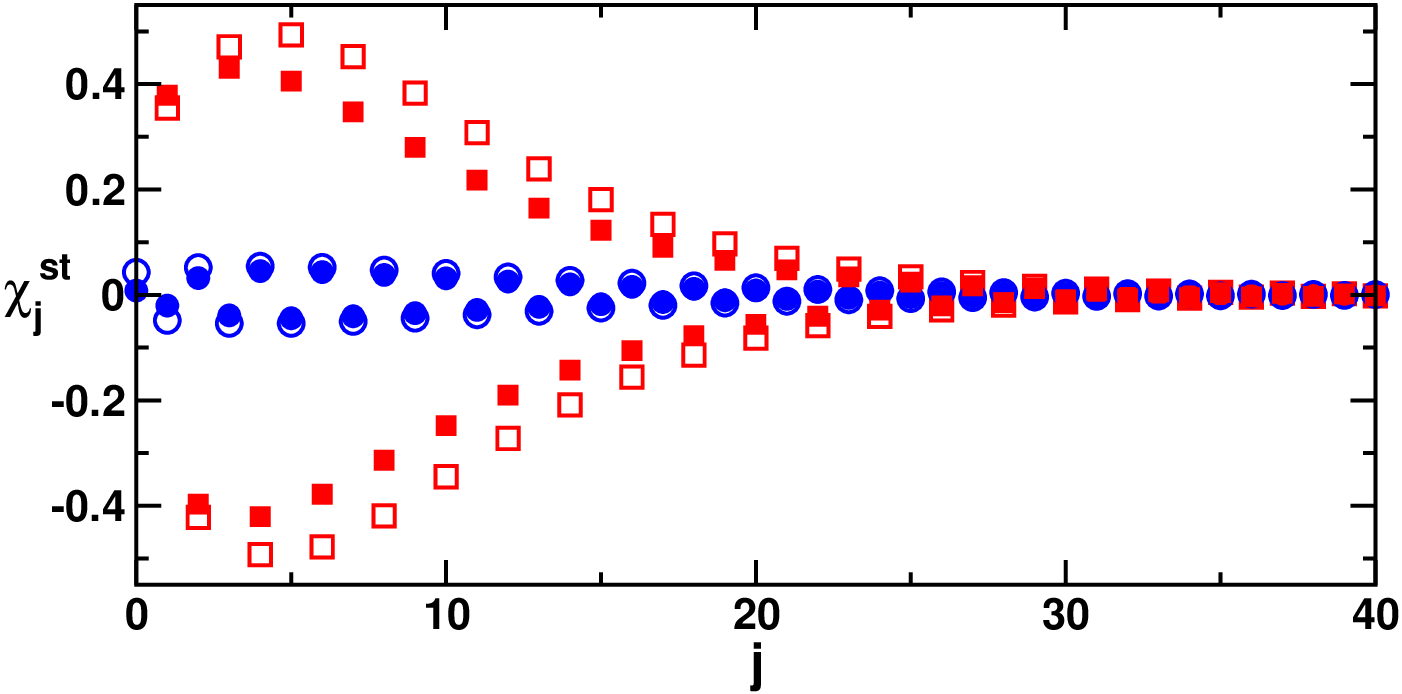}
\caption{(Color online) $\chi_j^{\rm st}$ for $\Delta=1$, $J_\perp =0.03J$,
  and $T=0.1J$. Comparison between DMRG data (closed symbols) and field theory
  (open symbols) for an infinite chain with a non-magnetic impurity at $j=0$
  (squares) and an infinite neighboring chain without impurities (circles).}
\label{fig5}
\end{figure}

Finally, the first order reflection $\chi_j^{\rm st (1)}$ of susceptibility
oscillations in a neighboring chain, shown in fig.~\ref{fig2}(c), is given in
first order perturbation theory in $J_{\perp}$ by
\begin{equation}
\label{reflection}
\chi_j^{\rm st (1)} = -\frac{J_\perp}{T}(-1)^j \sum_k \chi_k^{\rm st} G^{zz,
  st}_{j-k} \, .
\end{equation} 
Here eq.~(\ref{StaggPart4}) has to be used for $\chi_k^{\rm st}$ and $G^{zz,
  st}_{j-k} =\langle S^z_jS^z_k\rangle^{\rm st} = c^2/\l[\frac{v}{\pi
  T}\sinh(\frac{\pi T}{v}|j-k|)\r]^{K}$ is the staggered part of the bulk
two-point correlation function. While it would be extremely difficult to
obtain accurate numerical data for two weakly coupled chains at temperatures
$T/J\sim 0.01$ as considered in fig.~\ref{fig2}(c) we can easily check formula
(\ref{reflection}) at higher temperatures but still $T/J\ll 1$. Particularly
suited to study the case of an infinite chain with a single non-magnetic
impurity at the origin which is weakly coupled to an infinite chain without
impurities is the density-matrix renormalization group (DMRG) applied to
transfer matrices. This algorithm allows it to directly obtain results in the
thermodynamic limit \cite{SirkerEPL,SirkerBortz}. In fig.~\ref{fig5}, DMRG data
are compared to the field theoretical formulas
(\ref{StaggPart4},\ref{reflection}) and good agreement is found.
The maximum of $|\chi_j^{\rm st (1)}|$ and therefore the separation of the
shoulders $Dh$ scales like $T^{-3/2}$ for $K_L=1$ with complicated logarithmic
corrections coming in through the amplitude $c$. This might make it hard to
detect this power law in experiment.  The second order reflections can be
calculated analogously.
\section{Conclusions}
To conclude, we have derived an analytic formula for the local susceptibility
of a finite Heisenberg chain. This allows us to calculate NMR spectra for spin
chains with arbitrary impurity concentrations and distributions which would be
impossible by numerical calculations at temperatures $T/J \ll 1$. We also
showed how to calculate NMR spectra for weakly coupled spin chains if the
impurities are dilute. For SCO we have demonstrated excellent agreement
between our theory and experiment showing that SCO is indeed a prototypical
quasi one-dimensional spin chain compound. More generally speaking, we have
shown that NMR spectra are extremely useful to extract information about the
impurity concentration as well as about the magnetic couplings. In particular,
the coupling between the chains leads to an additional structure and its
position allows it to directly extract the coupling strength. We also want to
remark that this structure is very sensitive to the type of interchain
coupling. If two chains are coupled by a zigzag-interchain coupling, as is the
case, for example, for SrCuO$_2$ \cite{MotoyamaEisaki}, $\chi_j^{\rm st (1)}$
would be zero, i.e., there would not be any reflections to first order in
neighboring chains.  An analysis of NMR spectra can therefore also help to
clarify the geometry of the relevant magnetic exchange couplings. Furthermore,
we expect the results presented here to be also helpful for a more detailed
analysis of NMR spectra for systems like YBCO \cite{YamaniStatt} where CuO
chains and CuO$_2$ planes are weakly coupled.

\acknowledgments
The authors thank I.~Affleck and, in particular, S.~Eggert for helpful
discussions about the role of interchain couplings.

\end{document}